\documentclass[pra,twocolumn,showpacs,superscriptaddress,amsfonts,amsmath,
floatfix]{revtex4}
\usepackage{graphicx}
\usepackage{psfrag}%To include LaTeX subscripts on figures
\usepackage{epsfig}
\usepackage{epstopdf} % to include .eps graphics files with pdfLaTeX
\newcommand{\Journal}[4]{#1 {\bf #2}, #3 (#4)}
\newcommand{\PR}{Phys. Rev.}
\newcommand{\PRL}{Phys. Rev. Lett.}
\newcommand{\PRA}{Phys. Rev. A}
\newcommand{\PRB}{Phys. Rev. B}

\newcommand{\JMP}{J. Math. Phys.}

\begin{document}
\title {Phase diagram and momentum distribution of an interacting  Bose gas in a bichromatic lattice}
\author{Xiaolong Deng}
\affiliation{Universit\'e
Joseph Fourier, Laboratoire de Physique et Mod\'elisation des
Mileux Condens\'es, C.N.R.S. B.P. 166, 38042 Grenoble, France}

\author{R. Citro}
\email{citro@sa.infn.it} \affiliation{Universit\'e Joseph Fourier,
Laboratoire de Physique et Mod\'elisation des Mileux Condens\'es,
C.N.R.S. B.P. 166, 38042 Grenoble, France}
\affiliation{Dipartimento di Fisica E.R. Caianiello and
C.N.I.S.M., Universit\'a di Salerno, via S. Allende, 84081
Baronissi, Salerno, Italy}
\author{A. Minguzzi}
\email{anna.minguzzi@grenoble.cnrs.fr}
\affiliation{Universit\'e
Joseph Fourier, Laboratoire de Physique et Mod\'elisation des
Mileux Condens\'es, C.N.R.S. B.P. 166, 38042 Grenoble, France}
\author{E. Orignac}
\affiliation{Universit\'e de Lyon, Laboratoire de Physique de
l'\'Ecole Normale Sup\'erieure de Lyon, CNRS UMR5672, 46 All\'ee
d'Italie, 69364 Lyon Cedex 07, France}
\date{\today}

\begin{abstract}
We determine the phase diagram and the momentum distribution for a
one-dimensional Bose gas with repulsive short range interactions
in the presence of a two-color lattice potential, with
incommensurate ratio among the respective wave lengths, by using a
combined numerical (DMRG) and analytical (bosonization) analysis.
The system displays a delocalized (superfluid) phase at small
values of the intensity of the secondary lattice $V_2$ and a
localized (Bose glass-like) phase at larger intensity $V_2$. We
analyze the localization transition as a function of the height
$V_2$ beyond the known limits of free and hard-core bosons. We
find that weak repulsive interactions unfavor the localized phase
i. e. they increase the critical value of $V_2$ at which
localization occurs. In the case of integer filling of the primary
lattice, the phase diagram at fixed density displays, in addition
to a transition from a superfluid to a Bose glass phase, a
transition to a Mott-insulating state for not too large $V_2$ and
large repulsion.
We also analyze the emergence of a Bose-glass
phase
 by looking at the evolution of the Mott-insulator lobes when increasing
 $V_2$. The Mott lobes shrink and disappear
above a critical  value of $V_2$. Finally, we characterize the
superfluid phase by the momentum distribution, and show that  it
displays a power-law  decay at small momenta typical of Luttinger
liquids, with an exponent depending on the combined effect of the
interactions and of the secondary lattice. In addition, we observe
 two side peaks which
are due to the diffraction of the Bose gas by the second lattice.
 This latter feature
could  be observed in current experiments as characteristics of
 pseudo-random Bose systems.
\end{abstract}


\maketitle

\section{Introduction}

The interplay between disorder and interactions has been a
long-standing challenge for condensed matter theory. In the
absence of interactions a random potential can induce Anderson
localization \cite{Anderson}, i.e. make all the single-particle
eigenstates localized. In absence of disorder, bosons on a lattice
with repulsive interactions display, for commensurate filling,
 a superfluid (SF) to Mott insulator (MI) transition as
the repulsion is increased\cite{Fisher}, with the superfluid phase
displaying large density fluctuations and a gapless excitation
spectrum, while the Mott phase is incompressible and has a gap in
the excitation spectrum. If one considers both repulsive
interactions and disorder, these two effects will compete: while
disorder makes the bosons localized, short-range repulsive
interaction energy increases as the square of boson density and
hence the total energy of the system is minimized by depleting the
localized condensate towards a more uniform density distribution.
As a result, in a lattice Bose gas with short-range interactions a
novel Bose-glass (BG) phase, non superfluid yet compressible,
emerges between the superfluid and the
Mott-insulator\cite{Fisher}. From the experimental point of view,
it is possible to realize a system of bosons in a random potential
by placing ${}^4$He in porous media such as Vycor, aeorgels or
xerogels\cite{chan,wong}, or by using artificially disordered
Josephson junction networks\cite{mooij}. Experiments in porous
media revealed that the critical exponents of the
normal-superfluid transition in Helium were different from those
in pure helium in the case of aerogels and xerogels. However, the
aerogel and xerogel structures can hardly be described by a short
range correlated random potential. In the case of Josephson
junctions, localization of vortices was observed, but because of
dissipation, this system cannot be treated as fully coherent.  The
phase diagram of disordered boson system has also been
 intensively
studied by Quantum Monte Carlo
simulations\cite{Krauth,zhang,alet,hitchcock}. While some
conjectures made in Ref.\cite{Fisher} could be confirmed, it
appeared that very large system sizes were required to obtain
reliable results. Due to the theoretical difficulty of the
problem, one approach has been to reduce the spatial
dimensionality. In one dimension, it is known that in the absence
of interactions all states are localized as soon as the random
potential is non-zero \cite{lifshitz_book,mott_twose}. Moreover,
powerful specific techniques are available to handle the
interactions; this is the case e.g. of the bosonization
technique\cite{giamarchi_book_1d} or of the Density Matrix
Renormalization Group (DMRG) method\cite{White_DMRG,DMRG_review}.
For the specific case of a one-dimensional Bose gas subjected to
an uncorrelated disorder (in absence of a lattice), the phase
diagram has been obtained by Giamarchi and Schulz \cite{GiaSchu},
showing that while for zero interactions the system is always
localized, for nonzero values of the repulsive interactions a
superfluid phase is possible at small values of disorder.
Ref.\cite{GiaSchu} also predicted that  the non-superfluid
(Bose-glass) phase of an interacting Bose gas is expected to
differ markedly from the non-interacting Anderson-localized (AG)
phase, e.g. the density profile of a  Bose glass phase is rather
uniform, in contrast to the highly inhomogeneous density profile
of an ideal Bose gas in a disordered potential where all the
particles occupy the lowest single-particle localized orbital. The
phase diagram of a disordered, interacting Bose gas in
one-dimension has been the subject of several numerical
investigations by quantum Monte Carlo methods
\cite{Batrouni,Svistunov}, strong coupling
expansions\cite{monien96}, and Density-Matrix renormalization
group approaches\cite{DMRG_disorder}, that have established the
existence of a Mott insulating phase separated from the superfluid
phase by a Bose glass phase for disorder not excessively strong.
For stronger disorder, these numerical studies have established
that only the Bose glass and the superfluid is present. Also, the
existence of a superfluid dome in the phase diagram has been
obtained for the incommensurate case\cite{DMRG_disorder}.

With the development of atom cooling and trapping techniques,
studying the Mott transition of bosons has become experimentally
feasible \cite{bloch_MI_02}. Moreover, recent experiments with
ultracold atomic gases have realized a pseudo-disordered potential
by superimposing two optical lattices with incommensurate ratio
between  their spatial periodicities \cite{Fallani05} in a regime
where interactions are important. Experimentally it is possible to
characterize the system by measuring the excitation spectrum, the
momentum distribution and higher-order (e.g. noise) correlations
functions, as well as by looking at the equivalent of transport
behavior through the study of the damping of large-amplitude
dipole oscillations \cite{Fallani06}.

While the experiments performed with a bichromatic lattice were
focused on a regime where the lattice acts as a disorder
potential, the physics of a bichromatic lattice is much richer,
and the aim of this work is to describe the different possible
phases of an interacting Bose gas subjected to such lattices. In
the absence of interaction, the Schroedinger equation in a
bichromatic potential treated in the tight binding approximation
is known as the Harper model or the ``almost Mathieu problem'' and
has been extensively studied by solid state
physicists\cite{AubryAndre,hofstadter} and mathematical
physicists\cite{almostMathieu}. It is known to display a
delocalized regime for weak incommensurate potential, and a
localized regime for strong incommensurate potential, the two
regimes being related by a duality transformation. In the limit of
infinitely strong repulsions among the bosons (the so-called
Tonks-Girardeau regime), the problem can be solved by mapping to
an ideal spinless Fermi gas subjected to the same external
potential \cite{Girardeau1960}. In particular, the model displays
the same localization-delocalization threshold as in the
noninteracting case. However, the momentum distribution of the
Tonks-Girardeau bosons is not directly related to the one of the
spinless fermions, and for the specific case of the bichromatic
potential it has been studied in \cite{Rey}.  The case of spinless
fermions (or hard core bosons) with
 nearest neighbor repulsion was studied in
\cite{schuster}. We focus here on the regime of intermediate
repulsive interaction strengths.  In the case of commensurate
filling of the primary lattice and for $\Delta=0$ a Mott-insulator
phase is expected to occur at large values of interaction
strengths $U_c/t\simeq 3.3$ \cite{svistunov}. In the disordered
case, this Mott insulating phase competes with the localized
phase, and is expected to induce a Bose glass intermediate phase.

\begin{figure}
%[htbp]
%\centering
%\psfrag{x}{$n_0$}
%\psfrag{y}{$\gamma$}
{\includegraphics[width=8cm,angle=0]{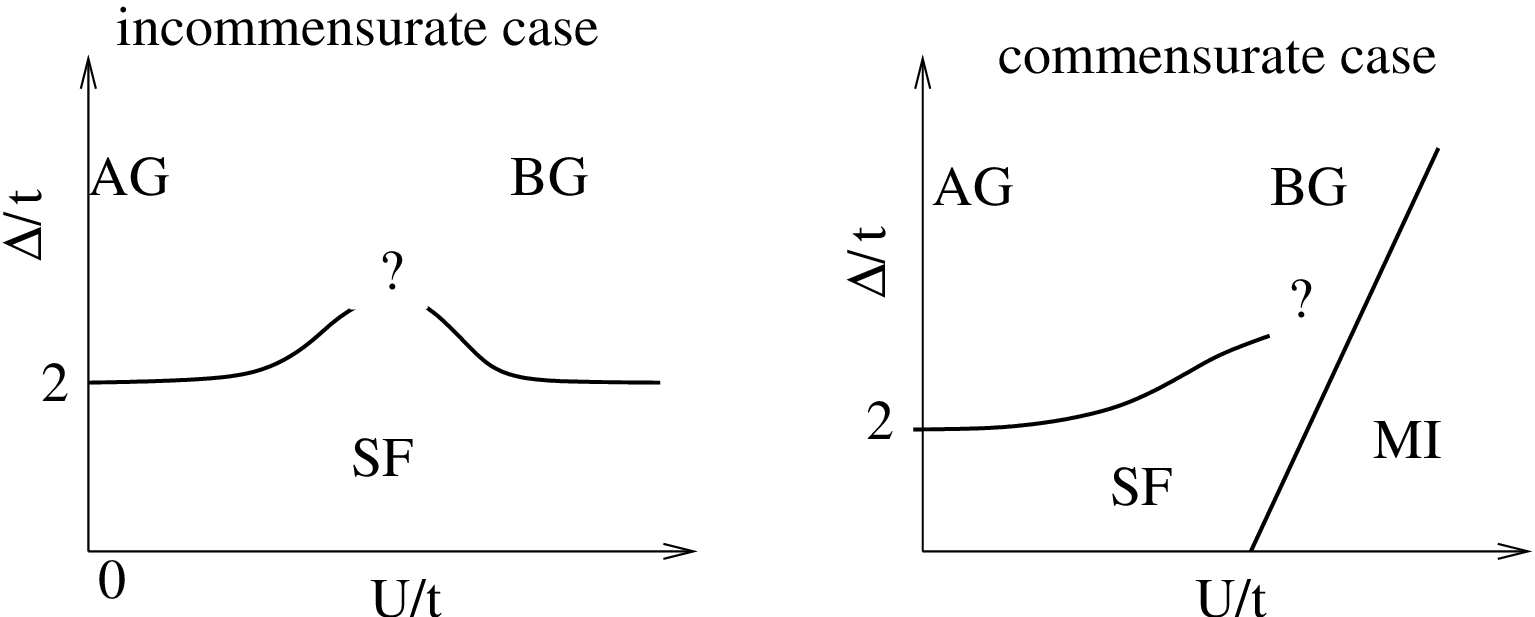}}
\caption{Schematic representation of the expected phase diagram
for a Bose gas subjected to a bichromatic potential. ``AG'' is the
Anderson-localized inhomogeneous phase, ``BG'' is the Bose glass
phase, ``SF'' is the non-localized superfluid-like phase (ie
displaying power-law decay of the phase-phase correlation
function), and ``MI'' is the Mott insulator phase.   The ``?''
sign stands for the region which need to be numerically
investigated.} \label{fig0}
\end{figure}

A Bose gas subjected to a quasiperiodic potential with of finite
interaction strengths has been previously studied by Roth et al.
\cite{Roth} by exact diagonalization on a very small system and by
Roscilde \cite{Roscilde08} in the case of a specific choice of the
height of the secondary lattice. In the present article, we use a
combination of density matrix renormalization group (DMRG) methods
and low-energy bosonization techniques to infer the phase diagram
of the gas at varying height of the secondary lattice and
interaction strengths, both for the case of integer and noninteger
filling of the main lattice.  The schematic summary of the known
limits of the phase diagram is presented in Fig.\ref{fig0}. One of
our aim is to see how the Mott lobes are modified by the presence
of the secondary lattice in the commensurate case and to establish
a phase diagram for both the commensurate and incommensurate case.
 We also  compute the momentum
distribution of the gas, which is one of observables
experimentally accessible.

The paper is organized as follows: In Sec.II we introduce the
model and the respective physical observables and give the
low-energy description of the system via bosonization approach.
Sec.III describes the numerical DMRG method. The results for the
phase diagram both for non-integer and integer filling at varying
the strength of the second lattice are given in Sec.IV. Here also
the evolution of the Mott-lobes with pseudo-disorder is given. In
Sec.V we analyze the momentum distribution function and describe
its characteristics for a weakly interacting Bose gas within
perturbation theory in the strength of the second lattice. In
Sec.V the dependence of the Luttinger exponent on pseudo-disorder
is also determined. Finally, in Sec.VI we give a summary and the
conclusions.

\section{Model}
We consider a one-dimensional Bose gas at zero temperature
subjected to a bichromatic lattice potential $V(x)=V_1 \sin^2(k_1
x) + V_2 \sin^2(k_2 x)$:

\begin{eqnarray}
\label{eq:cont_model} H=&&\int_{-\infty}^{\infty} dx \psi_b^\dagger
(x) \left( -\frac{\hbar^2}{2m} \nabla^2 +V(x)\right)\psi_b(x)\nonumber \\
&& +\frac g 2 \int _{-\infty}^{\infty} dx
\psi_b^\dagger(x)\psi_b^\dagger(x)\psi_b(x)\psi_b(x),
\end{eqnarray}
where $\psi_b(x)$ is the bosonic field operator, $m$ is the atomic
mass and $g$ represents the contact interaction. In the case where
the main lattice is quite large, i.e. $V_1\ge E_R$, where
$E_R=\hbar^2 k_1^2/2m$ is the recoil energy, we can map the system
on a Bose-Hubbard model\cite{jaksch}:

\begin{eqnarray}\label{eq:1} H &=&
-t\sum^{N_{sites}-1}_{i=1} (b^{\dagger}_i b_{i+1}+h.c.)+
\frac{U}{2}\sum^{N_{sites}}_{i=1} n_i(n_i -1)   \nonumber
\\&-&\mu\sum^{N_{sites}}_{i=1} n_i  +
\sum^{N_{sites}}_{i=1}\Delta_i n_i,
\end{eqnarray}
where $b^{\dagger}_i$, $b_i$ are bosonic field operators on the
site $i$, $t$ is the hopping amplitude, $U$ is the on-site
interaction, $\mu$ is the chemical potential, $N_{sites}$ is the
total number of lattice sites; the parameters $U,t$ are related to
those of the continuum model (\ref{eq:cont_model})(e.g. see Refs.
\cite{jaksch,Buchler03}). The effect of the second lattice is to
induce a modulation of the on-site energies according to
$\Delta_i=\Delta \cos(2 \pi \alpha i)$, with $\Delta\propto V_2$
is the relative strength of the second lattice, and  the value of
$\alpha=k_2/k_1$ \cite{note_irrational} has been chosen as $\alpha
= 830/1076 \approx 0.77$, being the same of the experiment in
Florence\cite{Fallani05}.

In order to characterize the different phases of the system, we evaluate the following observables: (i) the superfluid fraction,
\begin{eqnarray}
f_{s} = \frac{N_{sites}^2}{N
t\pi^2}\left(E^N_{antiPBC}-E^N_{PBC}\right),
\end{eqnarray}
where $N$ is the particle number, and
$E^N_{(anti)PBC}$ is the ground state energy with (anti)periodic
boundary conditions, and (ii) the compressibility, $\chi =(1/L)d
N/d\mu$, i.e.

\begin{equation}
\label{eq:compr} {\chi}^{-1}=L\lbrack E(N+1)+E(N-1)-2E(N)\rbrack,
\end{equation}
where $L$ is the length of the chain and $E$ the ground state
energy.

We also evaluate the momentum distribution as the Fourier transform of the one-body density matrix,
\begin{eqnarray}
n(q) = {\cal N} \sum_{lm}{\rm e}^{iq (l-m) a}\langle b^{\dagger}_l b_m\rangle.
\end{eqnarray}
with $a=\pi/k_1$ being the primary lattice spacing and $ {\cal N}$
a normalization constant.

\subsection{Low-energy properties and bosonization}

We focus now on the regime $\Delta\ll 2t$, which is expected to be
nonlocalized \cite{almostMathieu,AubryAndre,ZhongMos}. In these
conditions, we describe the one-dimensional interacting bosonic
fluid as a Luttinger liquid, using a low-energy hydrodynamic
description
\cite{haldane_bose_81,giamarchi_book_1d,cazalilla_review}. In
particular, the system is characterized by a slow, power-law decay
of the phase-phase correlation function (hence the denomination of
``superfluid phase'' ) with an exponent that depends on the
interaction parameters.

The low-energy  Hamiltonian for the fluid can be written
as~\cite{haldane_bose_81,giamarchi_book_1d}:
\begin{equation}
\label{eq:ham} H_0=\frac{1}{2\pi}\int dx \lbrack \frac{v_s}{K}
(\nabla \phi(x))^2 + v_s K (\pi \Pi(x))^2 \rbrack.
\end{equation}
This Hamiltonian is a standard sound wave one in which the
fluctuations of the phase $\phi(x)$ represent the phonon modes
of the density wave as given by
\begin{equation}\label{eq:density}
\rho(x)=\lbrack \rho_0-\frac 1 \pi \nabla \phi (x)\rbrack
\sum_{p=-\infty}^{\infty} e^{i2p(\pi \rho_0 x-\phi(x))},
\end{equation}
where $\rho_0$ is the average density of particles. The field
$\theta(x)=\pi \int^x dx' \Pi(x')$, is  conjugate of $\phi(x)$,
$[\frac 1 \pi \nabla \phi(x),\theta(x')]=i \delta(x-x')$ and
represents the phase of the superfluid. The parameters $K$ and
$v_s$  used in (\ref{eq:ham}) are related to the microscopic
compressibility and superfluid density  through the relations
$Kv_s= \pi \rho_s/m$ and $v_s/K=1/\pi\chi$. In the case of contact
interaction between bosons $g\delta(x)$ and in absence of the
lattice potential, the Luttinger parameters $v_s$ and $K$  are
obtained by the exact solution of the Lieb-Liniger model
~\cite{cazalilla_review}: $v_s K = \frac{\pi \rho_0}{m}$, as
follows from galilean invariance, and $\frac{v_s}{K} = \frac{g}{
\pi}$ in the weak coupling limit, while $\frac{v_s}{K} = \frac{\pi
\rho_0}{m}\left(1-\frac{8\rho_0 \hbar^2}{m g} \right)$ in the
strong coupling, $\rho_0=N/L$ being the particle density. The
Hamiltonian (\ref{eq:ham}) is an effective low-energy
theory\cite{haldane_bose_81} and provided that the correct values
of the parameters $v_s,K$ are used, all long wavelength properties
of the correlation functions of the system then can be obtained
exactly. In the $g=\infty$ limit, i.e. for hard-core bosons one
obtains $K=1$ as for free spinless fermions while the free bosons limit
would correspond to $K\rightarrow \infty$.

In the low-energy hydrodynamic description the bosonic field operator can be represented
as
\begin{equation}
\label{eq:bos_field} \psi_B(x)=e^{i\theta(x)}\sqrt{\rho(x)}.
\end{equation}
The corresponding one-body density matrix $G(x,x',0)=\langle
\psi_B^\dagger(x)\psi_B(x')\rangle$ in the long-wavelength limit
can be computed\cite{giamarchi_book_1d} and has a power-law decay
given by $\sim 1/|x-x'|^{\frac{1}{2K}}$ in the limit of the system
size $L\rightarrow  \infty$. Notice that the knowledge of the
compressibility and of the one-body density matrix offer two
independent ways of extracting the Luttinger exponent\cite{Monien}
$K$.

\subsection{Perturbative treatment of the quasiperiodic potential}

For the model (\ref{eq:ham}) we are interested in the effect of a
bichromatic lattice potential $V(x)=\sum_{i=1}^2 V_i \cos(2k_i
x)$. We will work in the limit where the strength of both potentials
are small with respect to the bandwidth, so that bosonization is
applicable.  Then,  each component $V_i$ of the potential couples to the
density and adds  a term to the Hamiltonian (\ref{eq:ham}) which
reads:
\begin{eqnarray}
\label{eq:bl} H_{bl}&=&V_i \int dx \cos(2 k_i x)\rho(x)\nonumber \\
&= &\sum_{p=-\infty}^{\infty} \frac{\rho_0 V_i}{2} \int dx
\cos\lbrack (2\pi p \rho_0\pm 2 k_i) x-2 p \phi(x) \rbrack.
\end{eqnarray}
Since the field $\phi(x)$ is a slowly varying  function on the
scale of the interparticle distance, if oscillating terms remain
in the integral, they will average out,  leading to a negligible
contribution. Therefore, the Luttinger liquid (superfluid)
behavior will persist provided that the filling is not
commensurate i.e. none of the two the commensurability conditions
$p \rho_0 \pm  k_i /\pi\in \mathbb{Z}$ are satisfied.

For commensurate fillings i. e. when one of the two commensurability
conditions is met,  the
periodic potential changes the simple quadratic Hamiltonian
(\ref{eq:ham}) of the Luttinger liquid into a sine-Gordon
Hamiltonian which describes  the Mott
transition as a function of  interaction strength~\cite{giamarchi_book_1d}.
Indeed, under the
renormalization-group (RG) flow, the operator (\ref{eq:bl}) is
irrelevant for $K>K_c=2/p^2$ and relevant for $K<K_c$,
thus implying a Mott-insulator phase at
$K<K_c$. As $K$ is decreasing when interactions are made more
repulsive, this means that the Mott state is obtained when repulsion
exceeds a critical value $U_c$. In the case where the Mott insulator
is obtained for $p=1$, in the regime of $K<K_c$, none of the terms
associated with the second potential (which is incommensurate) can
become relevant. Therefore, in that case, for $K>K_c$, the Luttinger
liquid is stable, and  no
Bose Glass phase can be created by the other potential in the
vicinity of the Mott Insulator superfluid transition in the regime
where bosonization is applicable. This justifies the shape of the
phase diagram of Fig.~\ref{fig0} for the commensurate case.
 The renomalization group analysis
shows that the transition from the Mott insulator to the superfluid
 belongs to the Kosterlitz-Thouless
universality class\cite{giamarchi_book_1d}.
 Note that the term
(\ref{eq:bl}) has been derived here for a weak lattice potential,
but it appears also in the opposite limit of a strong lattice
potential if the filling is commensurate, showing that the two
limits are smoothly connected\cite{giamarchi_book_1d}.

A different situation occurs in the case of random distributed
disorder. As shown in Refs.\cite{GiaSchu} the potential becomes
relevant below the critical value $K_c=3/2$. Below such value the
system lies in a Bose-glass phase with exponentially decaying
Green's function on the scale of the localization length.
A detailed RG analysis for the case of a generic quasiperiodic potential
 was given in Refs.\cite{vidal_qp_prl,vidal_qp_long}. There it was shown that
in the case where the quasiperiodic potential has a nontrivial, dense Fourier
spectrum, the critical value of $K_c$ can be actually
smaller than the value $K_c=2$, the
deviation from $K_c=2$ being related to the distance of $2\pi
\rho_0$ to a harmonic of the Fourier transform of the potential, thus
interpolating between the two-color potential and the fully random
case.

If we now consider the phase transition between the Mott state and
the superfluid, not as a function of interaction, but as a
function of particle density or as a function of the chemical
potential, it is well known that in the absence of the secondary
lattice potential, this is a commensurate-incommensurate (C-IC)
transition\cite{nersesyan_cic,giamarchi_book_1d,schulz_cic,Buchler03}.
At the transition, the scaling dimension of the operator $\cos
2\phi$ associated with the main lattice potential must be $1$,
which yields $K_c=1$. Turning on a second, weak lattice potential
incommensurate to the first, we see that the problem is reduced to
free fermions in a bichromatic lattice.
 The rigorous results on the
Harper model\cite{almostMathieu} then indicate that for a potential
which is small compared with the bandwidth, the states are not
localized by the incommensurate potential. Therefore, a weak
incommensurate potential cannot turn the superfluid state formed by
doping the Mott insulator in a Bose glass state. Again, this is at
variance with the effect of the random potential, which would
immediately localize the particles as the Mott gap closes.
 With
model~(\ref{eq:1}), in the
limit of very strong repulsion $U\gg t$, and for a filling slightly below
one particle per site, we can also use the Harper model mapping to
predict that the Bose glass to superfluid transition will happen when
$\Delta=2t$. Thus, in the phase diagram at fixed U, and varying $t$,
we expect that wings of Bose glass phase will be obtained for
sufficiently small $t$.
Summarizing the results for the Mott transition as a function of
chemical potential and interaction, we expect in the two-color
potential a scenario similar to the scenario 2(c) in \cite{Fisher},
i. e. that near the tip of the Mott lobe, there is no Bose glass phase
in the case of the two-color potential, provided that the
incommensurate potential is small compared to the bandwidth.

\section{Numerical method}

In order to determine the ground state properties of the
interacting Bose gas in the bichromatic lattice, we use the
Density Matrix Renormalization Group (DMRG)
method\cite{White_DMRG, DMRG_review}. The DMRG is a quasi-exact
numerical technique widely employed for studying strongly
correlated systems in low dimensions. Based on the
renormalization, it finds efficiently the ground state of a
relatively large system  with quite high precision. Recently, the
DMRG has already been used to study the 1D disordered Bose-Hubbard
model\cite{Rapsch_EPL}.

We consider a system with periodic boundary conditions and use
first the infinite-size algorithm to build the Hamiltonian up to
the length $L$, then we resort to the finite-size algorithm to
increase the precision within many sweeps. In principle the
Hilbert space of bosons is infinite; to keep a finite Hilbert
space in the calculation, we choose the maximal number of boson
states approximately of the order $5\langle n\rangle$, varying
$n_{max}$ between $n_{max}=6$ and $n_{max}=15$, except close to
the Anderson localization phase where we choose the maximal boson
states $n_{max}=N$. The number of eigenstates of the reduced
density matrix are chosen in the range $80-200$. To check the
error produced by truncating the boson space, we have repeated the
calculations at varying $n_{max}$ in the range $5\langle n\rangle$
and $10\langle n\rangle$, without observing substantial difference
in the ground state energy. To test the accuracy of our DMRG
method, in the case $U=0$ or for finite $U$ and small chain, we
have compared the DMRG numerical results with the exact solution
obtained by direct diagonalization. For larger system
($N_{sites}>10$), we have checked the convergence of the ground
state energy by varying the number of truncated eigenstates,
estimating that in the region of the superfluid-Mott insulating
phase the errors are of the order $10^{-6}$. The good convergence
of the algorithm is also tested by the coherence of the results
obtained from different observables as the Mott-insulator density
plateaus and correlation functions.

The calculations are performed in the canonical ensemble, i.e. at
fixed number of particles $N$. The chemical potential is
determined by the evaluation of the energy required to add or
subtract a particle to the ground state, i.e.
$\mu^{p}=E(N+1)-E(N)$ and $\mu^{h}=E(N)-E(N-1)$\cite{Batrouni}. In
this way we may obtain the phase diagram in the grand canonical
ensemble. In order to find the superfluid density and the
compressibility at varying chemical potential, we performed
several calculations at varying  particle numbers.  For the
determination of the phase diagram we have chosen $N_{sites}=20$,
while the correlation functions have been calculated using a
larger chain $N_{sites}=50$.

\section{Phase diagrams}

We have determined the phase diagram in two situations. First, we
have analyzed the effect of interactions on the
localization/delocalization threshold with respect to its
noninteracting value $\Delta=2 t$ obtained from the Harper model
\cite{AubryAndre} or equivalently for the hard-core Bose gas.
Secondly, we have analyzed the effect of disorder on the
Mott-insulator lobes \cite{Fisher}.

\subsection{Localization/delocalization transition}

\subsubsection{Incommensurate filling: case $\langle n\rangle=1/2$}

By the calculation in the canonical ensemble of the superfluid
fraction and of the compressibility, we have evaluated the phase
diagram in the plane ($\Delta/t$,$U/t$). This is illustrated in
Fig.~\ref{fig1} (upper panel) by showing the contour plot of the
superfluid fraction obtained for $N_{sites}=20$. In the case of
non-integer filling only two phases are present: a superfluid
phase ($f_s\ne 0$) at small values of the
secondary lattice height $\Delta$ (bottom-left), and a Bose glass
phase ($f_s=0$) at large values of $\Delta$ for
$\Delta>U$ (top-left). At $U=0$ the transition occurs at the
expected critical value $\Delta/t=2$. We see that at intermediate
values of the interaction strengths $U$ the critical value of
$\Delta_c/t$ increases and there the superfluid region extends in
a large dome. A similar behavior is observed for a disordered Bose
gas \cite{GiaSchu}.

\subsubsection{Commensurate filling: case $\langle n\rangle =1$}

\begin{figure}
%[htbp]
\centering
%\psfrag{x}{$n_0$}
%\psfrag{y}{$\gamma$}
{\includegraphics[width=7.5cm,angle=0]{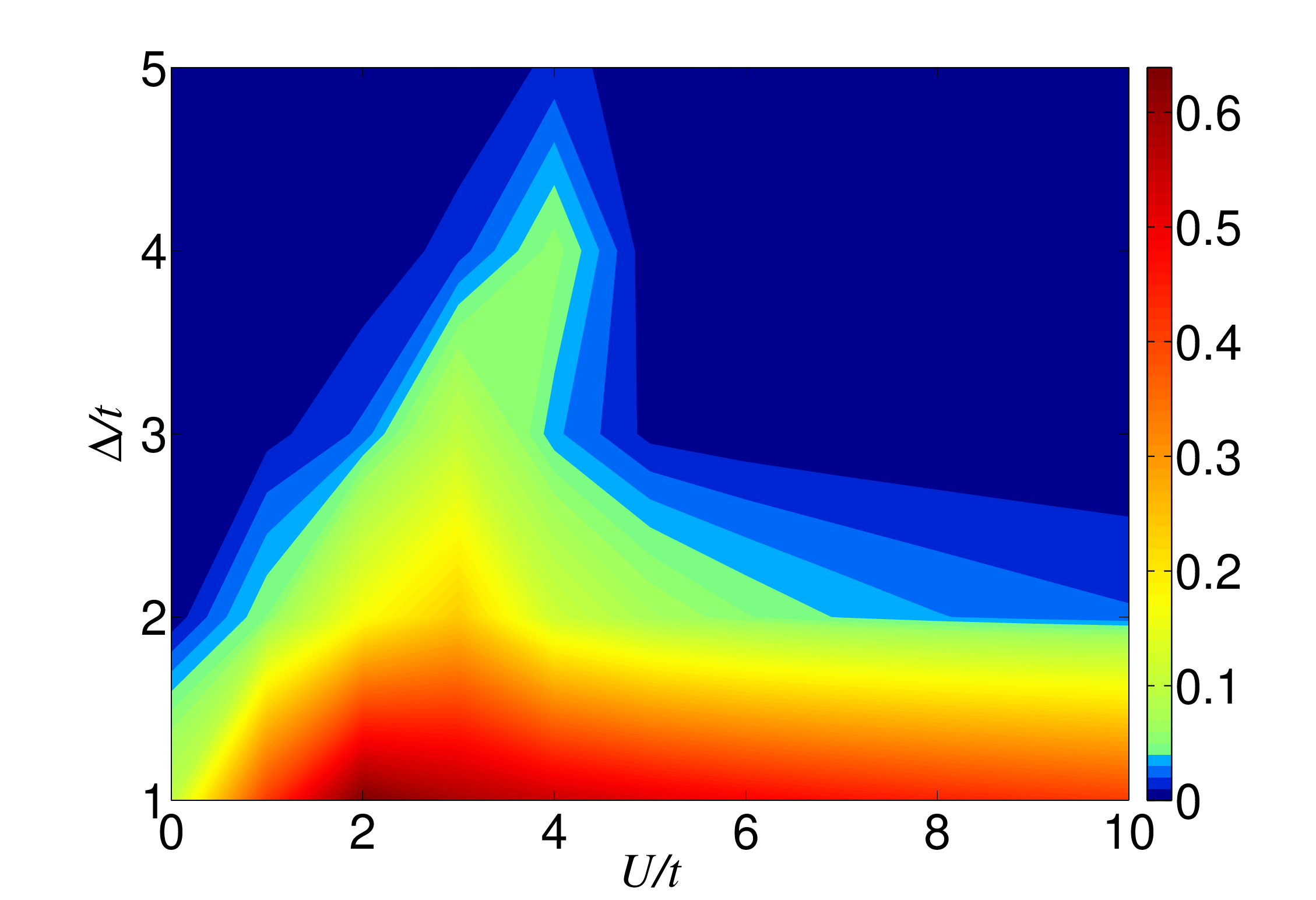}}\\
{\includegraphics[width=7.5cm,angle=0]{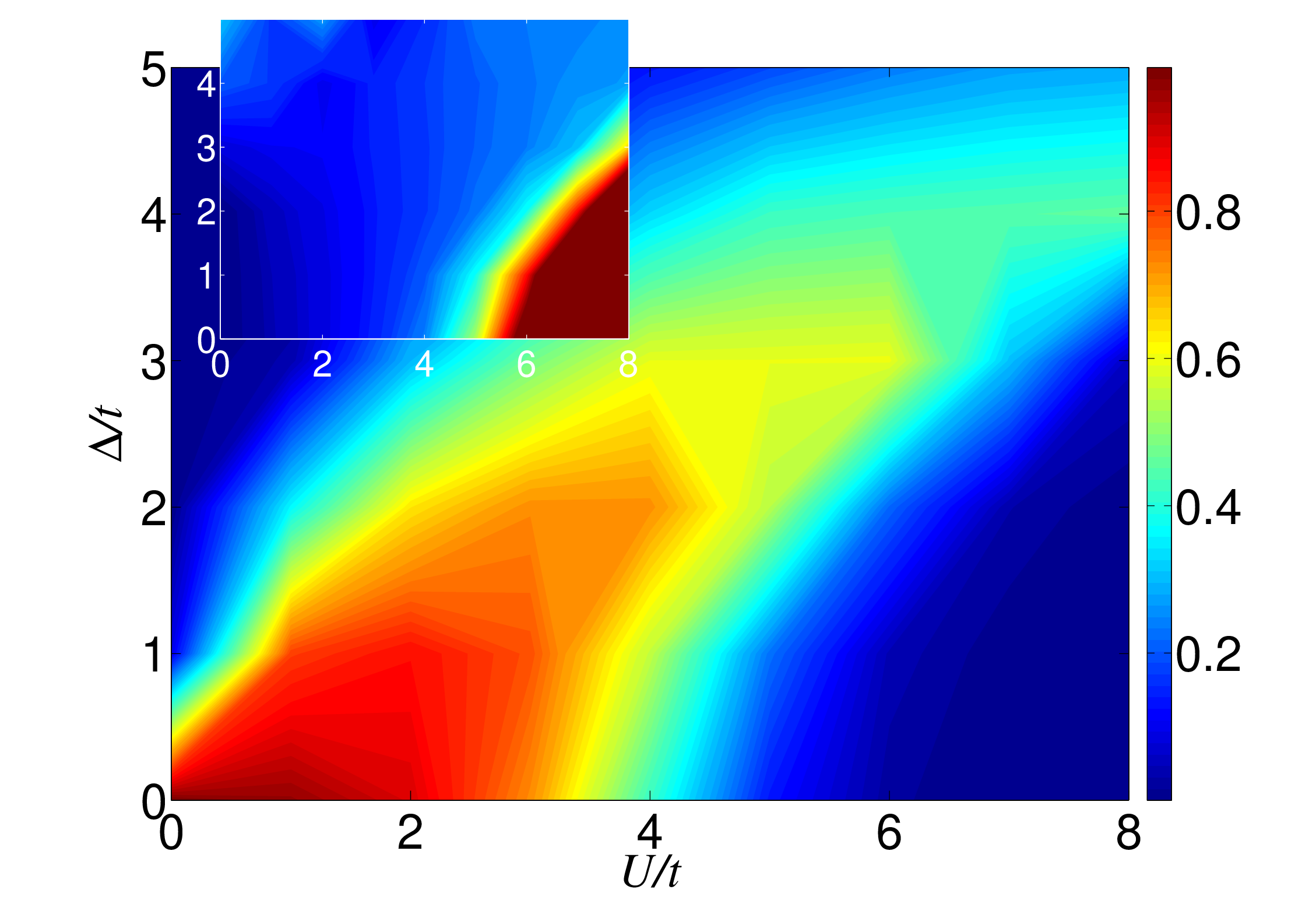}}
%{\includegraphics[width=6cm,angle=0]{onehalfphase.eps}}
\caption{DMRG phase diagram for an interacting Bose gas in a
two-color lattice, in the plane ($\Delta/t$ , $U/t$). Upper panel:
superfluid fraction in the case of non-integer filling
$\nu=N/N_{sites}=0.5$, with $N=10$, $N_{sites}=20$. Lower panel:
the superfluid fraction  $f_s$ (main figure) and compressibility gap $(\mu^p-\mu^h)/t$ (inset)  in the case of integer filling
with $N=N_{sites}=20$.}
\label{fig1}
\end{figure}

The phase diagram for the integer filling is given in
Fig.~\ref{fig1} (lower panel) where are reported the superfluid
fraction $f_s$ (main figure) and the compressibility gap $(\mu^p-\mu^h)/t$ (inset) obtained for $N_{sites}=20$.
 The Mott-phase which is characterized by a large
compressibility gap emerges at the bottom right corner above the
critical value $U_c/t=3.3\pm 0.2$ for $\Delta=0$ in agreement with
Ref.\cite{svistunov,Monien}. We observe that $U_c$ increases at
increasing $\Delta$, meaning that disorder energetically reduces
the compressibility gap in the localized regime (see also the upper panel of Fig.\ref{fig2} below).
A Bose-glass phase instead
occurs in the region of the phase diagram $\Delta>U$(top-left). At
$U=0$ the transition occurs at the expected value $\Delta_c/t=2$.
The critical value of $\Delta_c$ increases with $U$ at small $U$
indicating a delocalization by interactions, similarly to the true-disorder
case. Finally, a superfluid phase emerges in the small
$U$ and small $\Delta$ region of the phase diagram (bottom-left).
In our simulations it extends in a large dome at intermediate $U$
and $\Delta$. The behavior of the superfluid fraction and
compressibility gap for small $\Delta$ and intermediate $U$ seems
to indicate a direct transition from the
superfluid to the Mott-insulating state without passing into a
Bose glass. Such conclusion seems physically reasonable if one
takes into account that the bichromatic lattice potential acts as
a quasi-disorder, i.e. is less relevant than true disorder. Anyway
such conclusion should be supported by further numerical
investigation and finite size scaling of the compressibility
 and superfluid fraction.

\subsection{Mott-insulator lobes}

\begin{figure}[h]
%[htbp]
\centering
%\psfrag{x}{$n_0$}
%\psfrag{y}{$\gamma$}
{\includegraphics[width=6cm,angle=0]{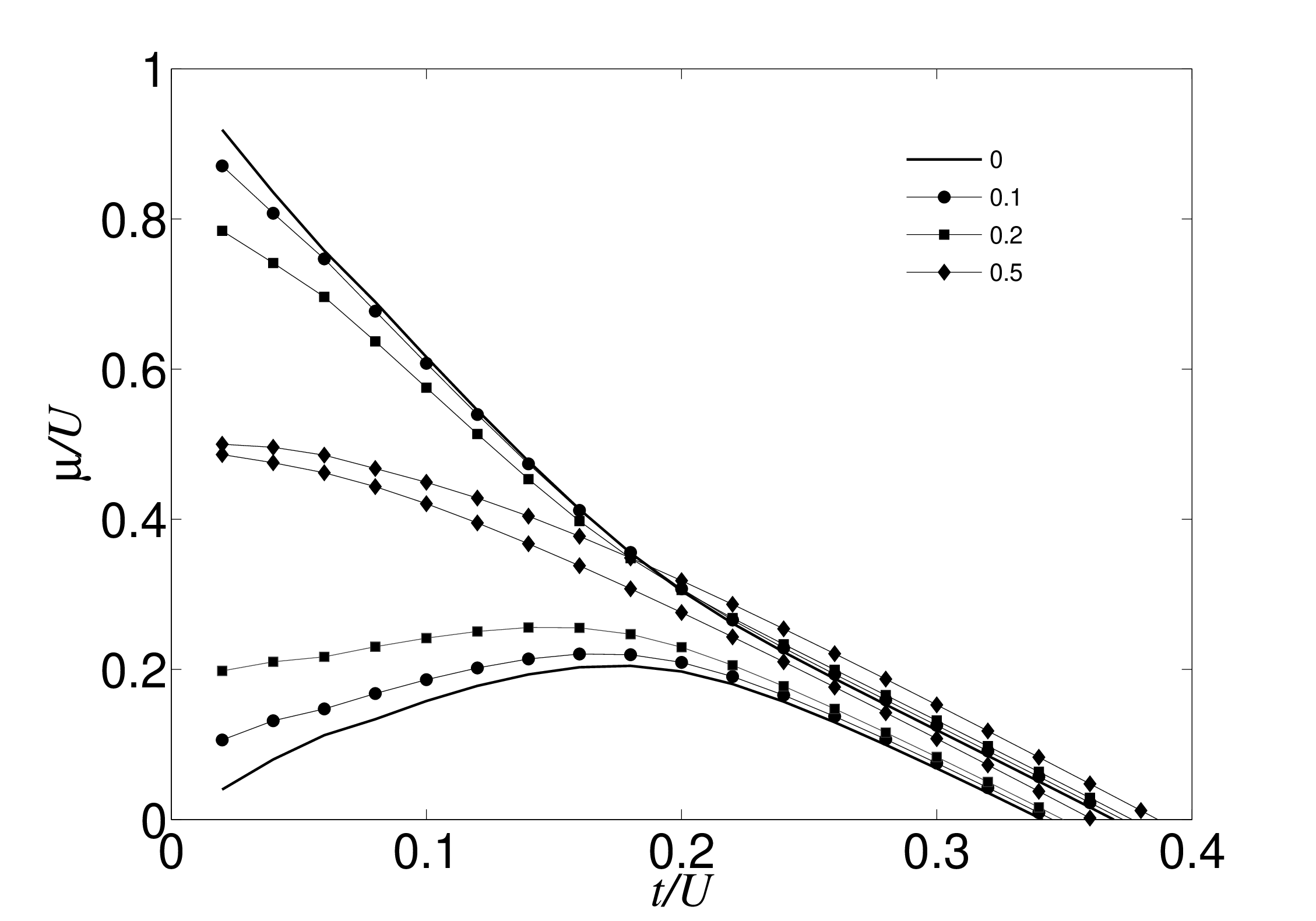}}
{\includegraphics[width=6cm,angle=0]{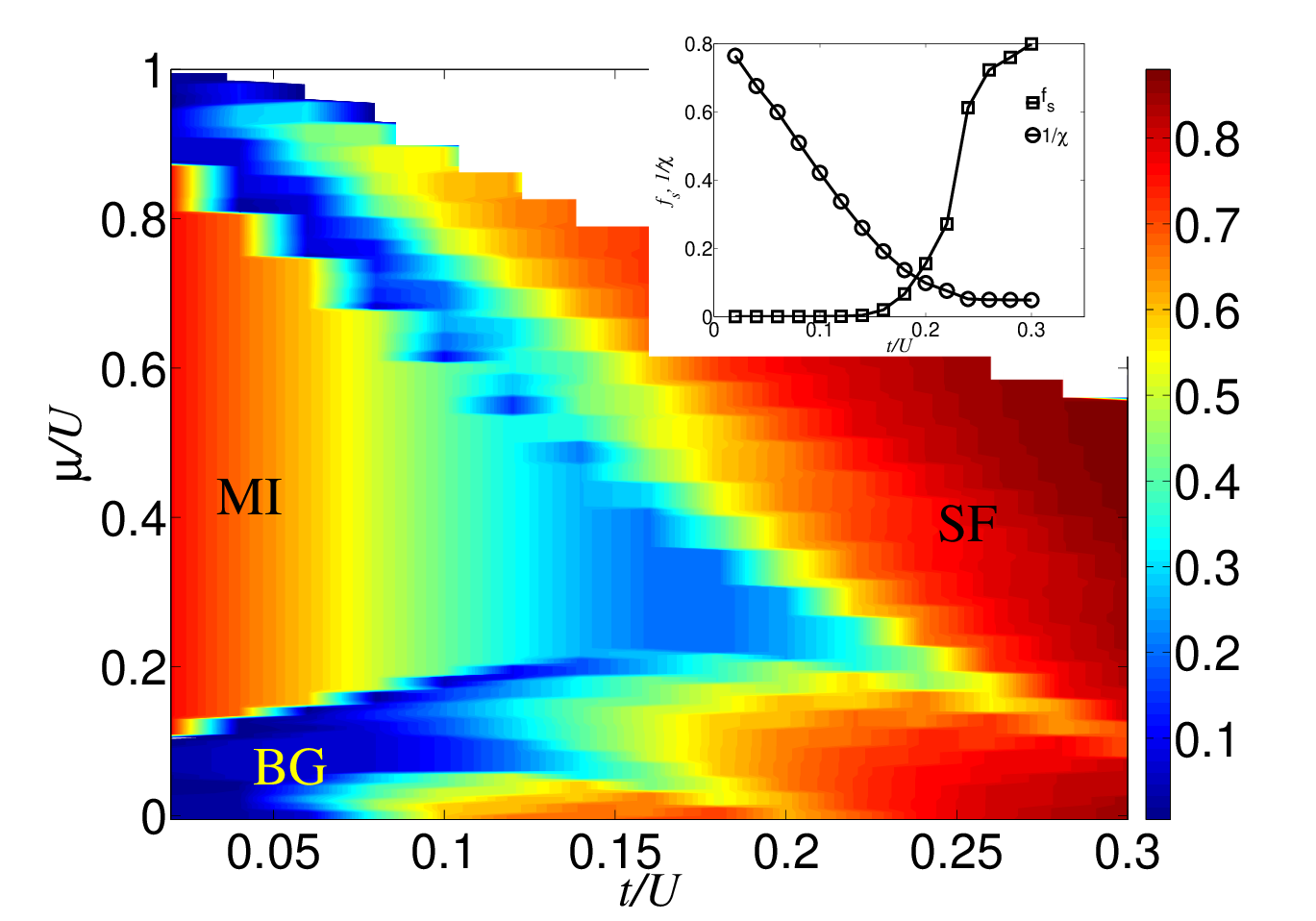}}
\caption{DMRG phase diagram for an interacting Bose gas in a two-color
  lattice, in the plane ($\mu/t$ , $t/U$), for the first Mott lobe,
  for $N=N_{sites}=20$. Upper panel: the shrinking of the Mott lobe at varying $\Delta/U$=0 (solid line), 0.1 (circles), 0.2 (squares), 0.5 (diamonds).
  Lower panel: contour plot of the function $f_s+(\mu^p-\mu^h)/t$ for
  $\Delta/U=0.1$. The inset shows the compressibility gap $(\mu^p-\mu^h)/t$ and the
  superfluid fraction $f_s$ along the line $\mu/U=0.25$.}
\label{fig2}
\end{figure}

We have performed the calculation of the Mott-insulator lobes in
the grand canonical ensemble. This is obtained by the estimation
of $\mu^p$ and $\mu^h$ for several values of particle numbers.
 At increasing strength of the second
lattice we find that the Mott-insulator lobe with $\langle
n\rangle=1$ shrinks and finally tends to disappear for $\Delta
\sim 0.5$, as is illustrated in Fig.~\ref{fig2} (upper panel). In
order to determine the Bose glass region we have also calculated
the superfluid density.  Fig.~\ref{fig2} (lower panel) shows, for
a specific choice of $\Delta$, the regions of nonzero superfluid
density as well as the regions of large  compressibility gap
(Mott-insulator phase) through the function $f_s+(\mu^p-\mu^h)/t$.
The intermediate (dark blue) region between the two corresponds to
the Bose glass phase. Notice that near the tip of the Mott lobe
the superfluid fraction is nonzero, as illustrated in the inset of
Fig.~\ref{fig2}(lower-panel), supporting the direct superfluid to
Mott-insulator transition, discussed above.  We also notice on
Fig.~\ref{fig2} the
presence of a Bose glass phase for $t/U\le \Delta/2U$, as expected
from the strong coupling argument.

\section{Momentum distribution}

\subsection{Side peaks of the momentum distribution}

\begin{figure}
%[htbp]
\centering
\psfrag{q}{$qa/\pi$}
\psfrag{n(q)}{$n(q)$}
{\includegraphics[width=6cm,angle=0]{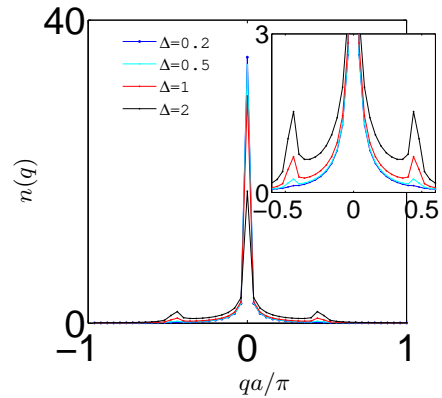}}
\caption{DMRG momentum distribution function in the superfluid
phase at varying $\Delta/U$ (as indicated on the figure) and $U=2 t$, with $N=N_{sites}=50$.
Subdominant peaks are determined by the presence of the second
lattice potential (see text).}
  \label{fig3}
\end{figure}

The results for the momentum distribution are reported in
Fig.\ref{fig3}. We note that besides the expected peak of the
momentum distribution at $k=0$, there are other peaks at $k=\pm
Q=\pm \frac{2\pi}{a}(1-\alpha)$ related to the modulation of the
on-site energy in Eq.(\ref{eq:1}). The origin of these peaks can
be understood by considering first non-interacting bosons. We will
begin by discussing the continuum limit, and then the lattice
case. If we approximate the irrational number $\alpha$
\cite{note_irrational} by a rational number $p/q$, in the
potential $V(x)$,  we can apply Bloch's theorem and write the
boson annihilation operator as:
\begin{eqnarray}
  \label{eq:boson}
  \hat{\psi_b}(x)=\frac{1}{\sqrt{N}} \sum_k \sum_{\beta=1}^q e^{i k x}
  \varphi_k^{(\beta)}(x) b_{k,\beta},
\end{eqnarray}
where $k$ is the quasi-momentum of the boson, and $\beta$ is the
band index. Bose condensation will then occur in the lowest
quasi-momentum state of the lowest band (we chose $\beta=1$ for
this band). The functions $ \varphi_k^{(\beta)}(x)$ are periodic
of period $qa$, i.e. $ \varphi_k^{(\beta)}(x)=
\varphi_k^{(\beta)}(x+qa)$.  Using this property one finds that in
the Bose Condensed state, $\langle \psi_b^\dagger(x+qa)
\psi_b(x'+qa) \rangle = \langle \psi_b^\dagger(x) \psi_b(x')
\rangle$. As a result, after averaging over $x$, the function
$\langle \psi_b^\dagger(x+r) \psi_b(x) \rangle$ becomes a periodic
function of $r$ of period $qa$. Using Fourier transformation, we
conclude that the states of momentum $(2\pi/a)(1/q\pm m)$ present a
macroscopic occupation number. If we turn to perturbation theory,
in the limit of $\Delta \ll t$, we find that the perturbed
wavefunction at the lowest order is given by
\begin{eqnarray}
\label{eq:pt}
 && \varphi_k^{(1),1}(x) =
 \varphi_k^{(1),0}(x) \\ \nonumber
 &&+\sum_{Q,m}\frac{
 \varphi_k^{(m),0}(x)}{E_{Q,m}-E_{k,1}}\langle \varphi_k^{(m),0}(x)|V_2 \cos(2 \alpha k_1
 x)|\varphi_k^{(1),0}(x)\rangle,
\end{eqnarray}
where $\varphi_k^{(m),0}(x)$ are the solutions of a Mathieu
equation\cite{Slater52} for the potential $V_1\cos(2k_1 x)$ and
$E_{Q,n}$ is the dispersion of the $n-$th band for momentum $Q$.
The matrix elements of perturbation are non zero only when
$Q=Q_{\pm}=(2 \pi/a) (\alpha \pm m)$ ($m \in \mathbb{Z})$.

The momentum distribution is then given by $n(p)=|\int dx e^{ipx}
\varphi_{k=0}^{(1),1}(x)|^2$ and using Eq.(\ref{eq:pt}) we find
that it displays two peaks:
\begin{equation}
\label{eq:mdan} n(p)\sim
|\varphi_0^{0}(p)|^2+\sum_{\delta=\pm}|\frac{V_2}{E_{Q_\delta}-E_0}|^2|\varphi_0^{0}(p+Q_\delta)|^2.
\end{equation}
where\cite{Slater52} $\varphi_0^{0}(p)\propto e^{-p^2/p_0^2}$ and
$p_0=\frac \pi a \left(\frac{E_R}{8V_1} \right)^{1/4}$.
%Considering a sequence of rational approximants converging to the
%irrational $\alpha$, we expect that the discrete spectrum of
%momentum states with macroscopic occupation number becomes a dense
%spectrum, with the dominant weight at $k=0$ and  subdominant
%weights $k=(2\pi/a)( \alpha\pm m)$.

In an analogous way we can proceed to derive the expression for the
momentum distribution on the lattice.
 The perturbed boson annihilation operator is then:
\begin{eqnarray}
\label{eq:op_pt}
  b_{k}=b_k^{(0)}+\sum_{\delta=\pm}\frac{\Delta}{-2t(\cos((k+Q_\delta)a)-\cos (ka))}
  b_{k+Q_\delta}^{(0)},
\end{eqnarray}
so that the largest occupation number will be found for $k=0$, and
again  at $k=Q_{\pm}$ (modulo the reciprocal lattice vector).
 The physical interpretation of the extra peaks is
therefore that the ground state wavefunction is diffracted by the
quasiperiodic potential thus creating peaks at multiple harmonics
of $2\pi \alpha/a$ (modulo a vector of the reciprocal lattice).

Let us now turn to the case of weakly interacting bosons. If the
repulsion $U$ is not too large, we can still begin by
diagonalizing the non-interacting Hamiltonian, and treat the
interaction within Bogoliubov approximation or numerically solve
the Gross-Pitaevski equation\cite{stringari_epjb}. Since Bose
condensation is obtained in the lowest band, it is reasonable to
neglect the contribution from the higher bands. Moreover, since
the states that are important for the low energy properties are
those with quasi-momentum near zero, we can neglect the dependence
of $\varphi_k^{(1)}(x_i)$ on $k$. This gives us the following
expression for the boson annihilation operator\cite{note}:
\begin{eqnarray}
  \label{eq:boso-approx}
  b_i \simeq \varphi_{0}^{(1)} (x_i) \tilde{b}_i,
\end{eqnarray}
where $\tilde{b}_i=\frac 1 {N^{1/2}} \sum_k e^{ik x_i} b_{k,1}$.
Injecting this approximation in the full Hamiltonian, we obtain an
interaction term which has the same period $q$ as the potential
$\Delta_i$. This gives rise to new umklapp processes, but since we
are only interested in the states of momenta close to zero, we can
neglect them. Then, the theory describing the $\tilde{b}$ bosons
becomes identical to the one describing bosons in the absence of
incommensurate potential, albeit with a dispersion fixed by the
band structure and an interaction $U_{eff.}=U\sum_{i=0}^{q-1}
|\varphi_{0}^{(1)} (x_i)|^4/q$.

The single particle density matrix is:
\begin{eqnarray}
  \langle b^\dagger_i b_j \rangle = (\varphi_{0}^{(1)} (x_i))^*
  \varphi_{0}^{(1)} (x_j)   \langle \tilde{b}^\dagger_i \tilde{b}_j
  \rangle,
\end{eqnarray}
and thus the effect of the periodic potential is only seen in the
appearance of the factor $(\varphi_{0}^{(1)}
(x_i))^*\varphi_{0}^{(1)} (x_j)$. Using the bosonization technique
to compute the single particle density matrix $\langle
\tilde{b}^\dagger_i \tilde{b}_j
  \rangle$, we finally find that:
  \begin{eqnarray}
\label{eq:rho1}
    \langle b^\dagger_i b_j \rangle = \frac{(\varphi_{0}^{(1)} (x_i))^*
  \varphi_{0}^{(1)} (x_j)}{|i-j|^{1/(2K)}}.
  \end{eqnarray}
By Fourier transforming the above expression, we recover power law
peaks in the momentum distribution with exponent $[1/2K-1]$ for
all the wavevectors that are multiples of $2\pi/qa$ . Based on the
previous perturbation theory, we expect that the two subleading
peaks will be found at $k=(2\pi/a)(m\pm p/q)$.
Moreover, the exponent should be identical to the one
found for $q=0$. We also remark that if the peaks were produced by
the terms $e^{i 2\pi
  \rho_0 x} e^{i (\theta -2\phi)}$ in the expansion of the boson
annihilation operator (\ref{eq:bos_field}), their position would
depend on the number of particle per site, and their height would
be independent of the strength of the incommensurate potential.
Moreover, these terms give in real space a correlation function of
the form $(|x-x'|/\alpha)^{-(2K+1/2K)}$ with an exponent that is
always larger than two. As a result, the Fourier transform of this
term would not diverge as $k \to (2\pi/a)(m\pm p/q)$,  than a cusp
could be obtained.

\begin{figure}
%[htbp]
\centering \psfrag{q}{$qa/\pi$} \psfrag{n(q)}{$n(q)$}
{\includegraphics[width=7cm,angle=0]{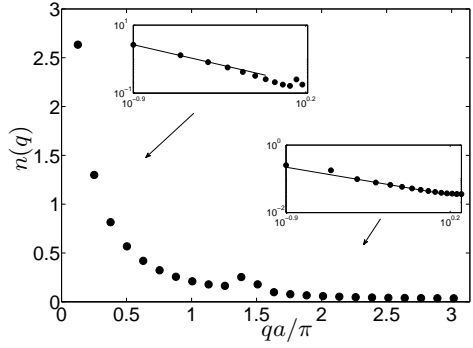}}
\caption{Fourier transform of DMRG momentum distribution function
in the superfluid phase with $\Delta=0.5 U$, $U=2 t$ and
$N=N_{sites}=50$. The main peak and the subdominant one  decay
with a power-law exponent consistent with $[1/2K-1]\sim 0.85$
for $q$ sufficiently close
to $0$ and $2\pi(1-\alpha)/a$ as shown in log scale in the
insets.} \label{fig:power-law}
\end{figure}

We have checked that the height of the secondary peak increases
quadratically with the strength of the incommensurate potential as
expected from (\ref{eq:mdan}), that its position does not change
with the filling, and that it possesses the same power law
divergence as the peak obtained at $k=0$. This is shown in
Fig.\ref{fig:power-law} where the Fourier transform of the
momentum distribution is displayed together with the power-law
decay of the peak at $q=0$ and of the satellite peak in a log
scale.

\subsection{Determination of the Luttinger exponent}

\begin{figure}
%[htbp]
\centering \psfrag{x}{$j$} \psfrag{rho1(x,0)}{$\rho_1(j,0)$}
{\includegraphics[width=7cm,angle=0]{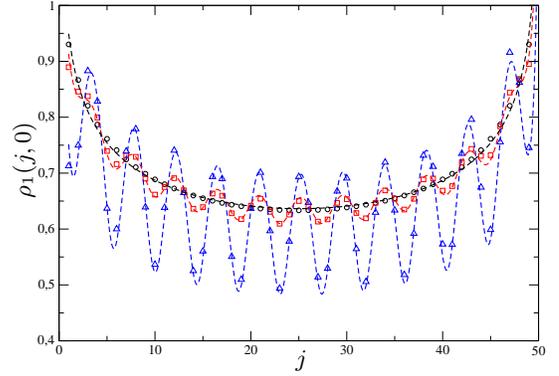}} \caption{One-body
density matrix  from DMRG data
($\Delta/U=0$ circles, $\Delta/U=0.1$ squares, $\Delta/U=0.5$
triangles) and from fit to Eq.(\ref{eq:fit}) (dashed lines). The
parameters used are $U=2t$ and $N=N_{sites}=50$.}
  \label{fig6}
\end{figure}

According to Eq.~(\ref{eq:rho1}), in the superfluid phase  the
one-body density matrix  $\rho_{1}(i,j)=\langle b^\dagger_i
b_j\rangle$ can be used to extract the Luttinger exponent $K$.
This is particularly interesting because, even though bosonization
techniques do not directly access to the localized phase, the fact
that the Luttinger exponent $K$ depends on the strength $\Delta$
of the  pseudo-disorder indicates a first disruption of the
superfluid phase towards localization.
 In order to analyze the DMRG data for the one-body density matrix,
we take into account both the density modulation induced by the
second lattice (entering explicitly in Eq.~(\ref{eq:rho1}) through
the factors $\varphi_0(x_i)$), and the fact that the calculations
are performed on a system of finite length $L$. For the latter
case, we use the results of the continuum model obtained by using
the conformal field theory \cite{cazalilla_review} for a system of
length $L$  and periodic boundary conditions. In essence, we fit the
DMRG results by the following expression:
\begin{equation}
\label{eq:fit}\rho_{1}(j,0)=\sqrt{n_0+ \delta \cos(2 \pi (1-\alpha) j+\phi_0)} \left[ \frac{1}{
d(j a|L)}\right]^{\frac{1}{2K}},
\end{equation}
where $n_0$, $\delta$ and $\phi_0$  are constants, K is the
Luttinger parameter and $d$ is the conformal length
$d(x|L)=\frac{L}{\pi}|\sin(\frac{\pi x}{L})|$. The results are
shown in Fig.\ref{fig6}. By the fit we obtain that the Luttinger
exponent $K$ decreases at increasing $\Delta$, in agreement with
the intuition that disorder drives the system towards a more
correlated, less superfluid phase. The corresponding values are
reported in Table~\ref{table}.

\begin{table}
\begin{tabular}{|c|c|}
  % after \\: \hline or \cline{col1-col2} \cline{col3-col4} ...
  \hline
  $\Delta/U$ & $K$ \\
  \hline
  0. &  3.44$\pm$ 0.03\\
  0.1 & 3.43$\pm$ 0.04 \\
  0.5 & 3.35$\pm$ 0.06 \\ \hline
\end{tabular}
\caption{Values of the Luttinger exponent from the fit  of the
DMRG data to Eq.(\ref{eq:fit}) with the parameters of
Fig.\ref{fig6}. The corresponding $\chi^2$ is of the order of
$5\times 10^{-5}$.} \label{table}
\end{table}

Another independent way to extract $K$ is based on the
determination of the ground state energy and compressibility
$\chi$ given by Eq.(\ref{eq:compr}), by the relation $K=\hbar \pi
\sqrt{ \rho_s \chi/m}$. We have verified that the values of $K$
extracted in this way are consistent with those of Table~\ref{table}.

\section{Summary and concluding remarks}

We have analyzed the phase diagram of an interacting
one-dimensional Bose gas in the presence of a pseudo-disorder generated by a
bichromatic lattice
potential. Starting
from a Bose-Hubbard model we have considered both commensurate and incommensurate fillings and we have found a rich phase diagram
including, in addition to the superfluid and Mott phases, a Bose
glass phase, localized but compressible. In agreement with the limiting cases
of free and hard-core bosons  described by an
almost Mathieu problem, the transition towards the Bose glass phase
is found at $\Delta/t\ge 2$, the critical value of $\Delta$ being  higher for
 bosons with finite interaction strength. This non-monotonic dependence of the critical  height of the second lattice on the
 interaction strength could be observed in the experiments.  We have also analyzed the shrinking of the Mott-lobes as a function of $\Delta$
and the emergence of a Bose-glass phase in the $(\mu/U,t/U)$
plane. Finally we have characterized the superfluid phase by a
static observable, the momentum distribution function. We have
shown that satellites peaks emerge when the pseudo-disorder is not
too strong and their interpretation within perturbation theory
offer a good qualitative understanding of their behavior as a
function of the height of the second lattice. The central peak of
the momentum distribution allows to determine the Luttinger
exponent $K$, whose knowledge is useful to make predictions for
further physical quantities.

While the momentum distribution and the behavior of the side peaks
could characterize the evolution of the system towards a Bose
glass, a direct probe of a Bose glass phase and its distinction
from a Mott-insulator could be provided by study of noise
correlations or collective excitations. This will be left for future
study.

\begin{acknowledgments}
R. Citro acknowledges financial support from a Marie-Curie
Intra-European Fellowship.  During the preparation of this
manuscript we have become aware of similar work by Roux et
al.\cite{Roux08}.
\end{acknowledgments}


\begin{thebibliography}{33}
\bibitem{note_irrational} In the definition of the Harper model
\cite{AubryAndre,almostMathieu} $\alpha$ is an irrational number. However,
to all practical purposes and in numerical calculations it is then approximated by a close rational number.
\bibitem{note} By following Ref.\cite{stringari_epjb} it can be
shown that the functions $\varphi_{0}^{(1)}$ satisfy the Bloch's
theorem and give rise to bands, thus an equation equivalent to
(\ref{eq:op_pt}) can be derived also for weakly interacting
bosons.
\end{thebibliography}
\end{document}